\documentclass[conference]{IEEEtran}

\usepackage{amsmath}
\usepackage{amssymb}
\usepackage[dvips]{graphicx}
\usepackage{epsfig}

\newcommand{\z}{\mathbf{z}}

\newcommand{\he}{\hat{h}}

\newcommand{\tsnr}{{\text{\footnotesize{SNR}}}}

\newtheorem{lemma:bitenergylow}{Lemma}
\newtheorem{lemma:bitenergyhigh}[lemma:bitenergylow]{Lemma}

\newtheorem{prop:asympcap}{Theorem}
\newtheorem{prop:flashminbitenergy}[prop:asympcap]{Theorem}
\newtheorem{prop:flashbitenergy}[prop:asympcap]{Theorem}
\newtheorem{prop:pasympcap}[prop:asympcap]{Theorem}

\begin{document}


\title{Achievable Rates and Optimal Resource Allocation for Imperfectly-Known Fading Relay Channels}

%\author{\authorblockN{Michael Shell} \and
%\and \authorblockN{James Kirk\\ and Montgomery Scott}
%\authorblockA{Starfleet Academy\\
%San Francisco, California 96678-2391\\ Telephone: (800)
%555--1212\\ Fax: (888) 555--1212}}

% avoiding spaces at the end of the author lines is not a problem with
% conference papers because we don't use \thanks or \IEEEmembership

\author{\authorblockN{Junwei Zhang, Mustafa Cenk Gursoy}
\authorblockA{Department of Electrical Engineering\\
University of Nebraska-Lincoln, Lincoln, NE 68588\\ Email:
jzhang13@bigred.unl.edu, gursoy@engr.unl.edu}}

%\specialpapernotice{(Invited Paper)}

\maketitle

\begin{abstract}\footnote{This work was supported in part by the NSF CARRER Grant CCF-0546384}
In this paper, achievable rates of imperfectly-known fading relay
channels are studied. It is assumed that communication starts with
the network training phase in which the receivers estimate the
fading coefficients of their respective channels. In the data
transmission phase, amplify-and-forward and decode-and-forward
relaying schemes are considered, and the corresponding achievable
rate expressions are obtained. The achievable rate expressions are
then employed to identify the optimal resource allocation
strategies.
\end{abstract}

\section{Introduction}

In wireless communications, deterioration in  performance is
experienced due to various impediments such as interference,
fluctuations in power due to reflections and attenuation, and
randomly-varying channel conditions caused by mobility and changing
environment. Recently, cooperative wireless communications has
attracted much interest as a technique that can mitigate these
degradations and have the performance approach to the levels
promised by multiple-antenna systems. Cooperative relay transmission
techniques have been studied in~\cite{jnl} and \cite{jnlbook} where
several two-user cooperative protocols have been proposed, with
amplify-and-forward~(AF) and decode-and-forward~(DF) being the two
basic modes. In~\cite{Nabar}, three different time-division AF and
DF cooperative protocols  with different the degrees of broadcasting
and receive collision are studied. In general, the area has seen an
explosive growth in the number of studies (see e.g.,
\cite{host-madsen1}, \cite{kramer}, \cite{Liang}, \cite{Mitran},
\cite{Yao} and references therein). However, most work has assumed
that the channel conditions are perfectly known at the receiver
and/or transmitter sides. Especially in mobile applications, this
assumption is unwarranted as the randomly-varying channel conditions
can be learned by the receivers only imperfectly. Recently, Wang
\emph{et al.} in \cite{wang} considered pilot-assisted transmission
over wireless sensory relay networks, and analyzed scaling laws
achieved by the amplify-and-forward scheme in the asymptotic regimes
of large nodes, large block length, and small $\tsnr$ values. In
this study, the channel conditions are being learned only by the
relay nodes.

In this paper, we study the achievable rates of imperfectly-known
fading relay channels. A priori unknown fading coefficients are
estimated at the receivers with the assistance of pilot symbols.
Following the training phase, AF and DF relaying techniques are
employed in the data transmission. Achievable rates for these
schemes are used to find the optimal resource allocation strategies.

\section{Channel Model}

We consider the three-node relay network which consists of a source,
destination, and a relay node. Source-destination, source-relay, and
relay-destination channels are modeled as  Rayleigh block-fading
channels with fading coefficients denoted by $h_{sr}$, $ h_{sd}$,
and $h_{rd}$, respectively, for each channel. Due to the
block-fading assumption, the fading coefficients $h_{sr}\sim\mathcal
{C}\mathcal {N}(0,{\sigma_{sr}}^2)$, $h_{sd}\sim\mathcal {C}\mathcal
{N}(0,{\sigma_{sd}}^2)$, and $h_{rd}\sim\mathcal {C}\mathcal
{N}(0,{\sigma_{rd}}^2)$ \footnote{$x\sim\mathcal {C}\mathcal
{N}(d,{\sigma^2)}$ is used to denote a proper complex Gaussian
random variable with mean $d$ and variance $\sigma^2$.} stay
constant for a block of $m$ symbols before they assume independent
realizations for the following block. In this system, the source
node tries to send information to the destination node with the help
of an intermediate relay node over the coherence block of $m$
symbols. The transmission is conducted in two phases: network
training phase and data transmission phase. Over these phases the
source and relay are subject to the following power constraints:
\begin{equation} \label{spower}
\|{\mathbf{x}_{s,t}}\|^2+E\{\|{\mathbf{x}_s}\|^2\}\leq mP_{s},
\end{equation}
\begin{equation} \label{rpower}
\|{\mathbf{x}_{r,t}}\|^2+E\{\|{\mathbf{x}_r}\|^2\}\leq mP_{r}.
\end{equation}
where $\mathbf{x}_{s,t}$ and $\mathbf{x}_{r,t}$ are the source and
relay training signal vectors respectively, and $\mathbf{x}_{s}$ and
$\mathbf{x}_{r}$ are the corresponding data transmission vectors.

\subsection{Network Training Phase}

Each block transmission starts with the training phase. In the first
symbol period, source transmits a pilot symbol to enable the relay
and destination to estimate channel coefficients $h_{sr}$ and
$h_{sd}$. In the average power limited case, sending a single pilot
is optimal because instead of increasing the number of pilot
symbols, a single pilot with higher power can be used. The signals
received by the relay and destination, respectively, are
\begin{equation} \label{srtraining}
y_{r,t}=h_{sr}x_{s,t}+n_r,
\end{equation}
\begin{equation} \label{sdtraining}
y_{d,t}=h_{sd}x_{s,t}+n_d.
\end{equation}
Similarly, in the second symbol period, relay transmits a pilot
symbol to enable the destination to estimate the channel coefficient
$h_{rd}$. The signal received by the destination is
\begin{equation} \label{rdtraining}
y_{d,t}^{r}=h_{rd}x_{r,t}+n_d.
\end{equation}
In the above formulations, $n_r\sim\mathcal {C}\mathcal {N}(0,N_0)$
and $n_d\sim\mathcal {C}\mathcal {N}(0,N_0)$ represent independent
Gaussian noise samples at the relay  and the destination nodes.

In the training process, it is assumed that the receivers employ
minimum mean-square error (MMSE) estimation. Let us assume that the
source allocates $\delta_s$ of its total power for training while
the relay allocates $\delta_r$ of its total power for training. As
described in  \cite{gursoy}, the MMSE estimate of $h_{sr}$ is given
by

\begin{equation}\label{est}
\hat{h}_{sr}=\frac{\sigma_{sr}^2\sqrt{\delta_smP_s}}{\sigma_{sr}^2\delta_smP_s+N_0}y_{r,t}
,
\end{equation}
where $y_{r,t}\sim\mathcal {C}\mathcal
{N}(0,\sigma_{sr}^2\delta_smP_s+N_0)$. We denote by
 $\tilde{h}_{sr}$  the estimate error which is a zero-mean complex
Gaussian random variable with variance
\begin{equation}
var(\tilde{h}_{sr})=\frac{\sigma_{sr}^2N_0}{\sigma_{sr}^2\delta_smP_s+N_0}.
\end{equation}
Similarly, we have
\begin{eqnarray}
\hat{h}_{sd}=\frac{\sigma_{sd}^2\sqrt{\delta_smP_s}}{\sigma_{sd}^2\delta_smP_s+N_0}y_{d,t},\nonumber\\
y_{d,t}\sim\mathcal {C}\mathcal
{N}(0,\sigma_{sd}^2\delta_smP_s+N_0),
\end{eqnarray}

\begin{equation}
var(\tilde{h}_{sd})=\frac{\sigma_{sd}^2N_0}{\sigma_{sd}^2\delta_smP_s+N_0},
\end{equation}
\begin{eqnarray}
\hat{h}_{rd}=\frac{\sigma_{rd}^2\sqrt{\delta_rmP_r}}{\sigma_{rd}^2\delta_rmP_r+N_0}y_{d,t}^r,\nonumber\\
y_{d,t}^r\sim\mathcal {C}\mathcal
{N}(0,\sigma_{rd}^2\delta_rmP_r+N_0),
\end{eqnarray}
\begin{equation}\label{est1}
var(\tilde{h}_{rd})=\frac{\sigma_{rd}^2N_0}{\sigma_{rd}^2\delta_rmP_r+N_0}.
\end{equation}
%%%%%%%%%%%%
With these estimates, the fading coefficients can now be expressed
as
\begin{equation}\label{ttt}
h_{sr}=\hat{h}_{sr} +\tilde{h}_{sr},
\end {equation}
\begin{equation}
 h_{sd}=\hat{h}_{sd}+\tilde{h}_{sd},
\end{equation}
\begin{equation}\label{tttt}
 h_{rd}=\hat{h}_{rd} +\tilde{h}_{rd}.
\end {equation}

\subsection {Data Transmission Phase}

The practical relay node usually cannot transmit and receive data
simultaneously. Thus, we assume that the relay works under
half-duplex constraint. We further assume that the relay operates in
time division duplex mode. As discussed in the previous section,
within a block of $m$ symbols, the first two symbols are allocated
for channel training. In the remaining duration of $m-2$ symbols,
data transmission takes place. First, the source transmits an
$(m-2)/2$-dimensional symbol vector $\mathbf{x}_s$ which is received
at the the relay and the destination, respectively, as
\begin{equation}\label{srtx}
\mathbf{y}_r=h_{sr}\mathbf{x}_{s}+\mathbf{n}_r,
\end{equation}
\begin{equation}\label{sdtx}
\mathbf{y}_d=h_{sd}\mathbf{x}_{s}+\mathbf{n}_d.
\end{equation}
Next, the source becomes silent, and the relay transmits an
$(m-2)/2$-dimensional symbol vector $\mathbf{x}_r$ which is
generated from the previously received $\mathbf{y}_r$ \cite{jnl}
\cite{jnlbook}. This approach  corresponds to protocol 2 in
\cite{Nabar}, which realizes the maximum degrees of broadcasting and
exhibits no receive collision. Thus, the destination receives
\begin{equation}\label{rdtx}
\mathbf{y}_d^r=h_{rd}\mathbf{x}_{r}+\mathbf{n}_d.
\end{equation}

After substituting (\ref{ttt})-(\ref{tttt}) into (\ref{srtx}),
 (\ref{sdtx}), (\ref{rdtx}) we have
\begin{equation}\label{srtx1}
\mathbf{y}_r=\hat{h}_{sr}\mathbf{x}_{s}+\tilde{h}_{sr}\mathbf{x}_{s}+\mathbf{n}_r,
\end{equation}
\begin{equation}\label{sdtx1}
\mathbf{y}_d=\hat{h}_{sd}\mathbf{x}_{s}+\tilde{h}_{sd}\mathbf{x}_{s}+\mathbf{n}_d,
\end{equation}
\begin{equation}\label{rdtx1}
\mathbf{y}_d^r=\hat{h}_{rd}\mathbf{x}_{r}+\tilde{h}_{rd}\mathbf{x}_{r}+\mathbf{n}_d.
\end{equation}
The input vectors $\mathbf{x}_{s}$ and $\mathbf{x}_{r}$ are assumed
to be composed of independent random variables with equal energy.
Hence the corresponding covariance matrices are
\begin{equation}\label{xsp}
E\{\mathbf{x}_{s}\mathbf{x}_{s}^\dagger\}=\frac{2(1-\delta_s)mP_s}{m-2}\mathbf{I},
\end{equation}
\begin{equation}\label{xrp}
E\{\mathbf{x}_{r}\mathbf{x}_{r}^\dagger\}=\frac{2(1-\delta_r)mP_r}{m-2}\mathbf{I},
\end{equation}
where $\mathbf{I}$ is the $(m-2)/2$-dimensional identity matrix.

\section{ A Capacity Lower-Bound For AF}

In this section, we consider the AF relaying scheme and calculate a
capacity lower bound using similar methods as those described in
\cite{training}. The capacity of the AF relay channel is the maximum
mutual information between the transmitted signal $\mathbf{x}_{s}$
and received signals $\mathbf{y}_d$ and $\mathbf{y}_d^r$ given $
\hat{h}_{sr}, \hat{h}_{sd}, \hat{h}_{rd}$. Thus, the capacity is
\begin{equation}
C=\sup_{p_{x_{s}}(\cdot )}\frac{1}{m}\emph{I}(\mathbf{x}_s
;\mathbf{y}_d,\mathbf{y}_d^r|\hat{h}_{sr},
 \hat{h}_{sd}, \hat{h}_{rd}).
\end{equation}
Note that this formulation presupposes that the destination has the
knowledge of $\hat{h}_{sr}$. Hence, we assume that the value of
$\hat{h}_{sr}$ is forwarded reliably from the relay to the
destination over low-rate control links.

Our method for finding a lower bound obtains $\hat{h}_{sr}$, $
\hat{h}_{sd}$ and $\hat{h}_{rd}$, relegates the estimation error of
channel estimates to the additive noise, and then considers only the
correlation (and not the full statistical dependence) between the
resulting noise and the transmitted signal. We then obtain a lower
bound by replacing the resulting noise by the worst case noise with
the same correlation. Let us assume that
\begin{equation}
\mathbf{z}_r=\tilde{h}_{sr}\mathbf{x}_{s}+\mathbf{n}_r,
\end{equation}
\begin{equation}
\mathbf{z}_d=\tilde{h}_{sd}\mathbf{x}_{s}+\mathbf{n}_d,
\end{equation}
\begin{equation}
\mathbf{z}_d^r=\tilde{h}_{rd}\mathbf{x}_{r}+\mathbf{n}_d,
\end{equation}
are the noise vectors which has the following covariance matrices:
\begin{equation}\label{zrp}
E\{\mathbf{z}_r\mathbf{z}_r^\dagger\}=\sigma_{\tilde{h}_{sr}}^2E\{\mathbf{x}_{s}\mathbf{x}_{s}^\dagger\}+N_0\mathbf{I},
\end{equation}
\begin{equation}\label{zdp}
E\{\mathbf{z}_d\mathbf{z}_d^\dagger\}=\sigma_{\tilde{h}_{sd}}^2E\{\mathbf{x}_{s}\mathbf{x}_{s}^\dagger\}+N_0\mathbf{I},
\end{equation}
\begin{equation}\label{zdrp}
E\{\mathbf{z}_d^r{\mathbf{z}_d^r}^\dagger\}=\sigma_{\tilde{h}_{rd}}^2E\{\mathbf{x}_{r}\mathbf{x}_{r}^\dagger\}+N_0\mathbf{I}.
\end{equation}
We therefore wish to find
\begin{multline}\label{ccc}
C\geqslant C_{worst}=\inf_{p_{z_{r}}(\cdot ),p_{z_{d}}(\cdot
),p_{z_{r}^d}(\cdot )}\\
\sup_{p_{x_{s}}(\cdot)}
\frac{1}{m}\emph{I}(\mathbf{x}_s;\mathbf{y}_d,\mathbf{y}_d^r|\hat{h}_{sr},
 \hat{h}_{sd},\hat{h}_{rd}).
\end{multline}

The following result provides $C_{worst}$.
\\
\begin{prop:asympcap} \label{prop:asympcap}
A lower bound on the capacity of AF scheme is given by
\begin{multline}\label{AFC1}
C_{worst}=\frac{m-2}{2m}E_{w_{sr}}E_{w_{sd}}E_{w_{rd}}\bigg[\log\Big(1+g(\delta_s,P_s,\sigma_{sd},w_{sd})\\+f\big[g(\delta_s,P_s,\sigma_{sr},w_{sr}),g(\delta_r,P_r,\sigma_{rd},w_{rd})\big]\Big)\bigg]
\end{multline}
where $w_{sr}\sim\mathcal {C}\mathcal {N}(0,1)$, $w_{sd}\sim\mathcal
{C}\mathcal {N}(0,1)$, $w_{rd}\sim\mathcal {C}\mathcal {N}(0,1)$,
and $f(x,y)=xy/(1+x+y)$. Furthermore $g(a,b,c,d)$ is defined as
{\small{
\begin{equation}\label{gg}
g(a,b,c,d)=\frac{2a(1-a)m^2b^2c^4}{2(1-a)mbc^2N_0+(m-2)(c^2amb+N_0)N_0}|d|^2.
\end{equation}}}
\end{prop:asympcap}

\emph{Proof}: For better illustration, we rewrite the channel
input-output relationships (\ref{srtx1}), (\ref{sdtx1}), and
(\ref{rdtx1}) for each symbol:
\begin{equation}\label{yri}
y_r[i]=\hat{h}_{sr}x_{s}[i]+z_r[i],
\end{equation}
\begin{equation}\label{ydi}
y_d[i]=\hat{h}_{sd}x_{s}[i]+z_d[i],
\end{equation}
for $i=3,4,...,(m-2)/2+2$, and
\begin{equation}\label{ydri}
y_d^r[i]=\hat{h}_{rd}x_{r}[i]+z_d^r[i],
\end{equation}
for $i=3+(m-2)/2,..., (m-2)+2$.\\
In AF, the signals received and transmitted by the relay have
following relation:
\begin{equation}
x_{r}[i]=\beta y_r[i-(m-2)/2],
\end{equation}
\begin{displaymath}
\beta\leqslant\sqrt{\frac{E[|x_r|^2]}{|\hat{h}_{sr}|^2E[|x_s|^2
]+E[|z_r|^2]}}.
\end{displaymath}
Now, we can write the channel in the vector form
\begin{eqnarray}\label{vec}
\underbrace{\left( \begin{array}{ccc}
y_d[i] \\
y_d^r[i+(m-2)/2]  \\
\end{array} \right)}_{\mathbf{\check{y}}_d[i]}
=\underbrace{\left( \begin{array}{ccc}
\hat{h}_{sd} \\
\hat{h}_{rd}\beta\hat{h}_{sr} \\
\end{array} \right)}_{A}x_s[i]+\nonumber\\
\underbrace{\left( \begin{array}{ccc}
0 &1&0 \\
\hat{h}_{rd}\beta&0&1 \\
\end{array} \right)}_{B}\underbrace{\left( \begin{array}{ccc}
z_r[i] \\
z_d[i]\\
z_d^r[i+(m-2)/2]
\end{array} \right)}_{\mathbf{z}[i]},
\end{eqnarray}
where $i=3,4,..., (m-2)/2+2$.
%For simplicity, we ignore the power of
%relay used for,\emph{ie.} in the separate control link or send extra
%symbols, to send estimated source-relay channel information
%$\he_{sr}$ to destination.We will see later that in DF we do not
%need to send this information to the destination.\\
With the  above notation, we can write the input-output mutual
information as
\begin{align}\label{ccccc}
\!\!\!\!I(\mathbf{x}_s;\mathbf{y}_d,\mathbf{y}_d^r|\hat{h}_{sr},
 \hat{h}_{sd},&\hat{h}_{rd})=\!\!\!\!\!\sum_{i = 3}^{(m-2)/2+2} \!\!\!\!\!\!I(x_s[i];\mathbf{\check{y}}_d[i]|\hat{h}_{sr},
 \hat{h}_{sd},\hat{h}_{rd})
 \\
 &\,\,\,\,\,\,=\frac{m-2}{2} I(x_s;\mathbf{\check{y}}_d|\hat{h}_{sr},
 \hat{h}_{sd},\hat{h}_{rd}) \label{eq:simplifiedmutualinfo}
\end{align}
where in (\ref{eq:simplifiedmutualinfo}) we removed the dependence
on $i$ without loss of generality. Note that $\mathbf{\check{y}}$ is
defined in (\ref{vec}). Now we can calculate the worst-case capacity
by proving that Gaussian distribution for $z_r$, $z_d$, and $z_d^r$
provides the worst case. Techniques similar to that
in~\cite{training} are employed. Any set of particular distributions
for $z_r$, $z_d$, and $z_d^r$ yields an upper bound on the worst
case. Let us choose $z_r$, $z_d$, and $z_d^r$ to be zero mean
complex Gaussian distributed. Then as in ~\cite{jnl},
\begin{align}\label{cworst1}
&\emph{C}_{worst} \le \frac{m-2}{2m} E \log
 \det\left(\mathbf{I}+(E(|x_s|^2)AA^\dagger)(BE[\z
 \z^\dagger]B^\dagger)^{-1}\right)
\end{align}
where the expectation is with respect to the fading estimates. To
obtain a lower bound, we compute the mutual information for the
channel (\ref{vec}), assuming that $x_s$ is a zero-mean complex
Gaussian with variance $E(|x_s|^2)$, but the distributions of noise
components $z_r$, $z_d$, and $z_d^r$ are arbitrary. Thus,
\begin{align}\label{I1}
&\emph{I}(x_s;\mathbf{\check{y}}_d;|\hat{h}_{sr},
\hat{h}_{sd},\hat{h}_{rd})
 =\emph{h}(x_s|\hat{h}_{sr},
 \hat{h}_{sd},\hat{h}_{rd})-\emph{h}(x_s|\mathbf{\check{y}}_d,\hat{h}_{sr},
 \hat{h}_{sd},\hat{h}_{rd})\nonumber\\
& \geqslant \log\pi e E(|x_s|^2)-\log\pi e \,
var(x_s|\mathbf{\check{y}}_d,\hat{h}_{sr},
 \hat{h}_{sd},\hat{h}_{rd}).
\end{align}
From \cite{training}, we know that
\begin{align}\label{cov}
var (x_s|\mathbf{\check{y}}_d,\hat{h}_{sr},
 \hat{h}_{sd},\hat{h}_{rd}) \leqslant
 E\left[(x_s-\hat{x}_{s})(x_s-\hat{x}_{s})^\dagger | \hat{h}_{sr},
 \hat{h}_{sd},\hat{h}_{rd}\right]
\end{align}
for any estimate $\hat{x}_{s}$ given
$\mathbf{\check{y}}_d,\hat{h}_{sr},
 \hat{h}_{sd}, \text{ and }\hat{h}_{rd}$. If we substitute the LMMSE estimate
$\hat{x}_{s}=R_{xy}R_{y}^{-1}\mathbf{\check{y}}_d$ into (\ref{I1})
and (\ref{cov}), we obtain \footnote{Here we use the property that
$\det(\mathbf{I}+\mathbf{A}\mathbf{B})=\det(\mathbf{I}+\mathbf{B}\mathbf{A})$}
\begin{align}
I(x_s;\mathbf{\check{y}}_d|\hat{h}_{sr},\hat{h}_{sd},\hat{h}_{rd})
\!\ge \!E \log
 \det\left(\mathbf{I}+(E[|x_s|^2]AA^\dagger)(BE[\z
 \z^\dagger]B^\dagger)^{-1}\right). \nonumber
\end{align}
As a result, we can easily see that
\begin{gather} \label{cworst2}
C_{worst}
 \geqslant \frac{m-2}{2m} E \log
 \det\left(\mathbf{I}+(E[|x_s|^2]AA^\dagger)(BE[\z
 \z^\dagger]B^\dagger)^{-1}\right).
\end{gather}
From (\ref{cworst1}) and (\ref{cworst2}), we have
\begin{gather} \label{cworst}
C_{worst} = \frac{m-2}{2m}E\log
 \det\left(\mathbf{I}+(E[|x_s|^2]AA^\dagger)(BE[\z
 \z^\dagger]B^\dagger)^{-1}\right).
\end{gather}
Now combining (\ref{ccc}), (\ref{ccccc})  and (\ref{cworst}), and
using the results (\ref{xsp}), (\ref{xrp}), (\ref{zrp}), (\ref{zdp})
and (\ref{zdrp}), we obtain the following capacity lower bound
\begin{eqnarray}\label{AFC}
C_{worst}=\frac{m-2}{2m}E_{\hat{h}_{sr}}E_{\hat{h}_{sd}}E_{\hat{h}_{rd}}\Bigg[\log\Bigg(1+\frac{\frac{2(1-\delta_s)mP_s}{m-2}{|\hat{h}_{sd}|}^2}{{\frac{2(1-\delta_s)mP_s}{m-2}}\sigma_{\tilde{h}_{sd}}^2+N_0}
\nonumber\\
+f\left(\frac{\frac{2(1-\delta_s)mP_s}{m-2}{|\hat{h}_{sr}|}^2}{{\frac{2(1-\delta_s)mP_s}{m-2}}\sigma_{\tilde{h}_{sr}}^2+N_0},\frac{\frac{2(1-\delta_r)mP_r}{m-2}{|\hat{h}_{rd}|}^2}{{\frac{2(1-\delta_r)mP_r}{m-2}}\sigma_{\tilde{h}_{rd}}^2+N_0}\right)\Bigg)\Bigg].
\end{eqnarray}
Intuitively, we may see the lower-bound as the case in which the
estimation error is completely detrimental.
%This result corresponds
%to the one in~\cite{medard}.
By substituting (\ref{est})-(\ref{est1})into (\ref{AFC}) and
normalizing, we can rewrite the capacity as in (\ref{AFC1})  \hfill
$\square$

\section{A Capacity Lower-bound for DF}
In DF, there usually  are two different coding approaches
\cite{jnlbook}, namely repetition coding and parallel channel
coding. We first consider repetition channel coding. For this case,
an achievable rate is
\begin{multline}\label{IDF1}
I_{RDF}=\frac{1}{m}\min\left\{\emph{I}(\mathbf{x}_s;\mathbf{y}_r|\hat{h}_{sr}),\emph{I}(\mathbf{x}_s;\mathbf{y}_d,\mathbf{y}_d^r|\hat{h}_{sd},\hat{h}_{rd})\right\}.
\end{multline}
Using this expression, we arrive to the following result.
\begin{prop:asympcap} \label{prop:asympcap1}
An achievable rate expression for DF with repetition channel coding
is given by
\begin{equation}\label{DFC1}
I_{worst}=\min\{I_1,I_2\}
\end{equation}
where
\begin{multline}\label{C1}
I_1=\frac{m-2}{2m}E_{w_{sr}}\bigg[\log\Big(1+g(\delta_s,P_s,\sigma_{sr},w_{sr})\Big)\bigg],
\end{multline}

\begin{multline}\label{C2}
I_2=\frac{m-2}{2m}E_{w_{sd}}E_{w_{rd}}\bigg[\log\Big(1+g(\delta_s,P_s,\sigma_{sd},w_{sd})\\+g(\delta_r,P_r,\sigma_{rd},w_{rd})\Big)\bigg]
\end{multline}
where $g(.)$ is defined in (\ref{gg}).
\end{prop:asympcap}

\emph{Proof}: As described in \cite{training}, we can obtain the
worst-case mutual information for the first term in (\ref{IDF1}) by
proving that Gaussian distributed $z_r$ is the worst case. This
gives us $I_1$. In repetition coding, after successfully decoding
the source information, the relay transmits the same codeword as the
source. As a result, we can rewrite the data transmission with
regard to the second mutual information in (\ref{IDF1})  as
\begin{eqnarray}\label{vec1}
\underbrace{\left( \begin{array}{ccc}
y_d[i] \\
y_d^r[i+(m-2)/2]  \\
\end{array} \right)}_{\mathbf{y}_d[i]}
=\underbrace{\left( \begin{array}{ccc}
\hat{h}_{sd} \\
\hat{h}_{rd}\beta \\
\end{array} \right)}_{A}x_s[i]+\nonumber\\
\underbrace{\left(
\begin{array}{ccc}
z_d[i]\\
z_d^r[i+(m-2)/2]
\end{array} \right)}_{z[i]}.
\end{eqnarray}
In repetition coding
\begin{equation}
\beta=\sqrt{\frac{E[|x_r|^2]}{E[|x_s|^2]}}.
\end{equation}
From (\ref{vec1}), it is clear that the knowledge of $\he_{sr}$ is
not required at the destination. We can easily see that (\ref{vec1})
is a simpler expression than what we have in the AF case, therefore
we can adopt the same methods as described in Section 3 to show that
Gaussian noise is the worst case which gives $I_2$. The resulting
lower bound capacity is
\begin{equation}\label{DFC1}
I_{worst}= \min\{I_1,I_2\},
\end{equation}
where
\begin{equation}\label{DFC11}
I_1=\frac{m-2}{2m}E_{\hat{h}_{sr}}\Bigg[\log\Bigg(1+\frac{\frac{2(1-\delta_s)mP_s}{m-2}{|\hat{h}_{sr}|}^2}{{\frac{2(1-\delta_s)mP_s}{m-2}}\sigma_{\tilde{h}_{sr}}^2+N_0}\Bigg)\Bigg],
\end{equation}
\begin{multline}\label{DFC21}
I_2=\frac{m-2}{2m}E_{\hat{h}_{sd}}E_{\hat{h}_{rd}}\Bigg[\log\Bigg(1+\frac{\frac{2(1-\delta_s)mP_s}{m-2}{|\hat{h}_{sd}|}^2}{{\frac{2(1-\delta_s)mP_s}{m-2}}\sigma_{\tilde{h}_{sd}}^2+N_0}\\
+\frac{\frac{2(1-\delta_r)mP_r}{m-2}{|\hat{h}_{rd}|}^2}{{\frac{2(1-\delta_r)mP_r}{m-2}}\sigma_{\tilde{h}_{rd}}^2+N_0}\Bigg)\Bigg].
\end{multline}
Again by substituting (\ref{est})-(\ref{est1}) into
(\ref{DFC1})-(\ref{DFC21}) and normalizing, we obtain Theorem
\ref{prop:asympcap1}. \hfill $\square$

If parallel channel coding is employed \cite{jnlbook}, then we have,
\begin{gather}\label{IDF}
I_{PDF}=\frac{1}{m}\min\left\{\emph{I}(\mathbf{x}_s;\mathbf{y}_r|\hat{h}_{sr}),\emph{I}(\mathbf{x}_s;\mathbf{y}_d|\hat{h}_{sd})
+\emph{I}(\mathbf{x}_r;\mathbf{y}_d^r|\hat{h}_{rd})\right\}.
\end{gather}
Again it can easily be shown that the worst case is experienced when
$z_r,z_d$, and $z_d^r$ are Gaussian distributed. The resulting
achievable rate is given in the following result.
\begin{prop:asympcap} \label{prop:asympcap2}
An achievable rate expression for DF with parallel channel coding is
\begin{equation}\label{DFC}
I_{worst}=\min\{I_1,I_2\},
\end{equation}
where
\begin{multline}\label{C11}
I_1=\frac{m-2}{2m}E_{w_{sr}}\bigg[\log\Big(1+g(\delta_s,P_s,\sigma_{sr},w_{sr})\Big)\bigg],
\end{multline}

\begin{multline}\label{C21}
I_2=\frac{m-2}{2m}E_{w_{sd}}E_{w_{rd}}\bigg[\log\Big(1+g(\delta_s,P_s,\sigma_{sd},w_{sd})\Big)\\+\log\Big(1+g(\delta_r,P_r,\sigma_{rd},w_{rd})\Big)\bigg]
\end{multline}
where $g(.)$ is defined in (\ref{gg}).
\end{prop:asympcap} \hfill $\square$

\section{Optimal Resource Allocation}

\begin{figure}
\begin{center}
\includegraphics[width = 0.45\textwidth]{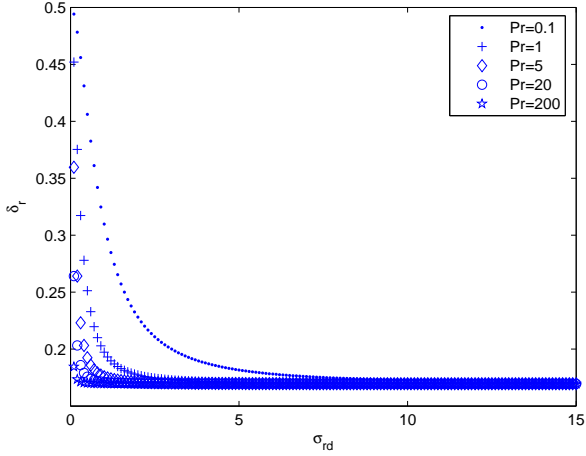}
\caption{$\delta_r$ vs. $\sigma_{rd}$ for different values of $P_r$
when $m =50$.} \label{fig:1}
\end{center}
\end{figure}

We first study how much power should be allocated for channel
training. In AF, it can be seen that $\delta_r$ appears only in
$g(\delta_r,P_r,\sigma_{rd},w_{rd})$ in the achievable rate
expression (\ref{AFC1}). Since $ f(x,y)=\frac{xy}{1+x+y} $ is a
monotonically increasing function of $y$ for fixed $x$, (\ref{AFC1})
is maximized by maximizing $g(\delta_r,P_r,\sigma_{rd},w_{rd})$. We
can maximize $ g(\delta_r,P_r,\sigma_{rd},w_{rd}) $ by maximizing
the coefficient of the random variable $|w_{rd}|^2$, and the optimal
$\delta_r$ is given by the expression in (\ref{taor}).
\begin{figure*}
\begin{equation}\label{taor}
\delta_r^{opt}=\frac{1}{2}\frac{-4mP_r\sigma_{rd}^2-2mN_0+4N_0+2\sqrt{-4m^2P_r^2\sigma_{rd}^4-2m^2P_r\sigma_{rd}^2N_0+m^2N_0^2-4mN_0^2+4N_0^2+2m^3P_r^2\sigma_{rd}^4+m^3P_r\sigma_{rd}^2N_0}}{-4mP_r\sigma_{rd}^2+m^2P_r\sigma_{rd}^2}.
\end{equation}
\end{figure*}
Optimizing $\delta_s$ is more complicated as it is related to all
the terms in (\ref{AFC1}), and hence obtaining an analytical
solution is unlikely. A suboptimal solution is to maximize
$g(\delta_s,P_s,\sigma_{sd},w_{sd})$ and
$g(\delta_s,P_s,\sigma_{sr},w_{sr})$ seperately, and obtain two
solutions $\delta_{s,1}^{subopt}$ and $\delta_{s,2}^{subopt}$,
respectively. Note that expressions for $\delta_{s,1}^{subopt}$ and
$\delta_{s,2}^{subopt}$ are exactly the same as that in (\ref{taor})
with $P_r$ and $\sigma_{rd}$ replaced by $P_s$, and $\sigma_{sd}$
and $\sigma_{sr}$, respectively.
%The first term is maximized by maximizing the coefficient of
%$|w_{sd}|^2$. The resulting $\delta_s$ is shown in (\ref{taos1})
%\begin{figure*}
%\begin{equation}\label{taos1}
%\delta_s^{best1}=\frac{1}{2}\frac{-4mP_s\sigma_{sd}^2-2mN_0+4N_0+2\sqrt{-4m^2P_s^2\sigma_{sd}^4-2m^2P_s\sigma_{sd}^2N_0+m^2N_0^2-4mN_0^2+4N_0^2+2m^3P_s^2\sigma_{sd}^4+m^3P_s\sigma_{sd}^2N_0}}{-4mP_s\sigma_{sd}^2+m^2P_s\sigma_{sd}^2}.
%\end{equation}
%\end{figure*}
%The second term  function is maximized by maximizing the coefficient
%of $|w_{sr}|^2$. The resulting $\delta_s$ is shown in (\ref{taos2})
%\begin{figure*}
%\begin{equation}\label{taos2}
%\delta_s^{best2}=\frac{1}{2}\frac{-4mP_s\sigma_{sr}^2-2mN_0+4N_0+2\sqrt{-4m^2P_s^2\sigma_{sr}^4-2m^2P_s\sigma_{sr}^2N_0+m^2N_0^2-4mN_0^2+4N_0^2+2m^3P_s^2\sigma_{sr}^4+m^3P_s\sigma_{sr}^2N_0}}{-4mP_s\sigma_{sr}^2+m^2P_s\sigma_{sr}^2}.
%\end{equation}
%\end{figure*}
When the source-relay channel is better than the source-destination
channel, $g(\delta_s,P_s,\sigma_{sr},w_{sr})$ is a more dominant
factor and $\delta_{s,2}^{subopt}$ is a good choice for training
power allocation. Otherwise, $\delta_{s,1}^{subopt}$ might be
preferred. For DF, similar results and discussions apply. For
instance, the optimal $\delta_r$ has the same expression as that in
(\ref{taor}).
%For $\delta_s$ things is a little more complex
% by maximize $I_1$ we get the optimal $\delta_s's$ expression is the
% same as shown in (\ref{taos2})
% by maximize $I_2$ we get the optimal $\delta_s's$ expression is the
% same as shown in (\ref{taos1}). When $I_1 \leq I_2$, the achievable
% rate is $I_1$, thus we should choose the $\delta_s^{best2}$. On the
% other hand, $\delta_s^{best1}$ should be chosen.
Figure \ref{fig:1} plots the optimal $\delta_r$ as a function of
$\sigma_{rd}$ for different relay power constraints $P_r$ when $m =
50$. It is observed in all cases that the allocated training power
decreases and convereges to a certain value with improving channel
quality.

\begin{figure}
\begin{center}
\includegraphics[width = 0.43\textwidth]{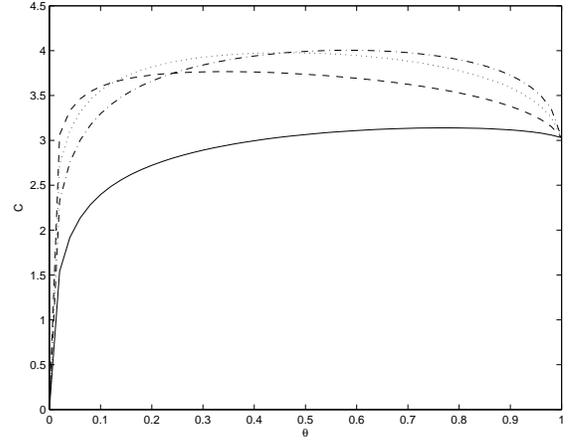}
\caption{AF achievable rate vs. $\theta$. $P =100$. (1) the dashed
line $\sigma_{sd}=1, \sigma_{sr}=10, \sigma_{rd}=2$; (2) the dotted
line $\sigma_{sd}=1, \sigma_{sr}=6, \sigma_{rd}=3$; (3) the dashdot
line $\sigma_{sd}=1, \sigma_{sr}=4, \sigma_{rd}=4$;(4) the solid
line $\sigma_{sd}=1, \sigma_{sr}=2, \sigma_{rd}=1$}. \label{fig:2}
\end{center}
\end{figure}

In certain cases, source and relay are subject to a total power
constraint. Here, we introduce the power allocation coefficient
$\theta$, and total power constraint $P$. $P_s$ and $P_r$ have the
following relations: $P_s=\theta P$, $P_r=(1-\theta)P$, and $P_s+P_r
= P$. Next, we investigate how different values of $\theta$, and
hence different power allocation strategies, affect the achievable
rates. An analytical result for $\theta$ that maximizes the
achievable rates is difficult to obtain. Therefore, we resort to
numerical analysis. First, we consider the AF. The parameters we
choose are $P=100, N_0=1, \delta_s=0.1, \delta_r=0.1$.
Fig.\ref{fig:2} plots the capacity lower bound (\ref{AFC1}) as a
function of $\theta$ for different channel conditions, i.e.,
different values of $\sigma_{sr},\sigma_{rd}, \text{ and
}\sigma_{sd}$. We observe that the best performance is achieved when
$\theta \approx 0.6$ and $\sigma_{sd}=1, \sigma_{sr}=4,
\sigma_{rd}=4$ which indicates that both source-relay and
relay-destination channels are favorable. When $\sigma_{sd}=1,
\sigma_{sr}=2, \sigma_{rd}=1$, and hence the relay-destination and
source-relay channels are not much better than the
source-destination channel, the optimal value of $\theta$ is close
to 1 and there is only little to be gained with cooperation.

\begin{figure}
\begin{center}
\includegraphics[width = 0.43\textwidth]{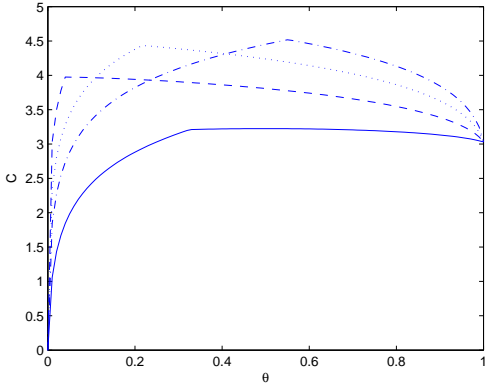}
\caption{Repetition coding  DF rate vs. $\theta$. $P =100$. (1) the
dashed line $\sigma_{sd}=1, \sigma_{sr}=10, \sigma_{rd}=2$; (2) the
dotted line $\sigma_{sd}=1, \sigma_{sr}=6, \sigma_{rd}=3$; (3) the
dashdot line $\sigma_{sd}=1, \sigma_{sr}=4, \sigma_{rd}=4$; (4) the
solid line $\sigma_{sd}=1, \sigma_{sr}=2, \sigma_{rd}=1$. %(1) and
%(4) are partly overlapped because they have the same s-d and r-d
%channel conditions.
} \label{fig:3}
\end{center}
\end{figure}

\begin{figure}
\begin{center}
\includegraphics[width = 0.40\textwidth]{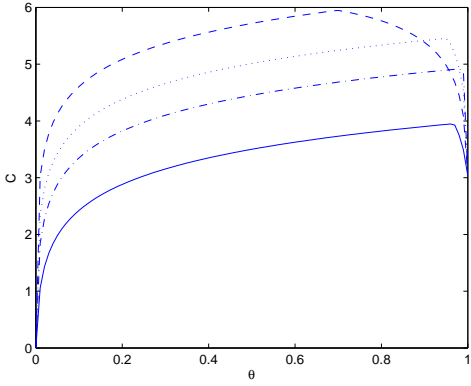}
\caption{Parallel coding DF rate vs. $\theta$. $P =100$. From the
top to bottom, the four curves correspond to (1) $\sigma_{sd}=1,
\sigma_{sr}=10, \sigma_{rd}=2$ ; (2) $\sigma_{sd}=1, \sigma_{sr}=6,
\sigma_{rd}=3$; (3) $\sigma_{sd}=1, \sigma_{sr}=4, \sigma_{rd}=4$;
(4) $\sigma_{sd}=1, \sigma_{sr}=2, \sigma_{rd}=1$. } \label{fig:4}
\end{center}
\end{figure}

Figs. \ref{fig:3} and \ref{fig:4} plot the DF achievable rates as a
function of $\theta$ with the same parameters as in the AF case.
Hence, the total power is $P = 100$. It is seen that paralel coding
achieves a better performance compared to that of repetition coding.
In parallel coding DF, we observe that unless $\sigma_{sr}$ is high
and hence the source-relay channel is strong, the optimal value of
$\theta$ is close to 1 and relay is allocated small power.

\begin{figure}
\begin{center}
\includegraphics[width = 0.40\textwidth]{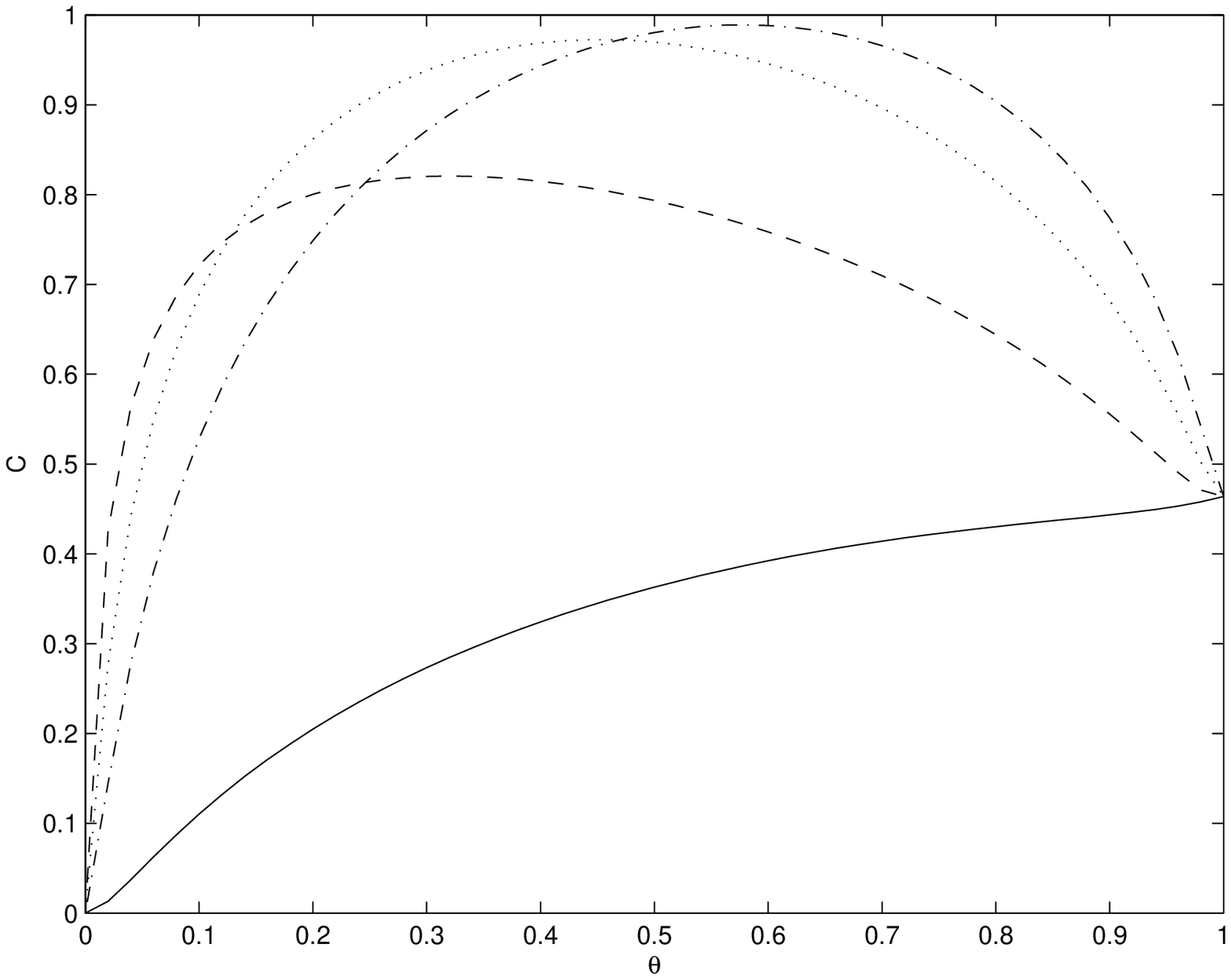}
\caption{AF achievable rate vs. $\theta$. $P =1$. 1) the dashed line
$\sigma_{sd}=1, \sigma_{sr}=10, \sigma_{rd}=2$; (2) the dotted line
$\sigma_{sd}=1, \sigma_{sr}=6, \sigma_{rd}=3$; (3)the dashdot line
$\sigma_{sd}=1, \sigma_{sr}=4, \sigma_{rd}=4$;(4) the solid line
$\sigma_{sd}=1, \sigma_{sr}=2, \sigma_{rd}=1$}. \label{fig:5}
\end{center}
\end{figure}

\begin{figure}
\begin{center}
\includegraphics[width = 0.43\textwidth]{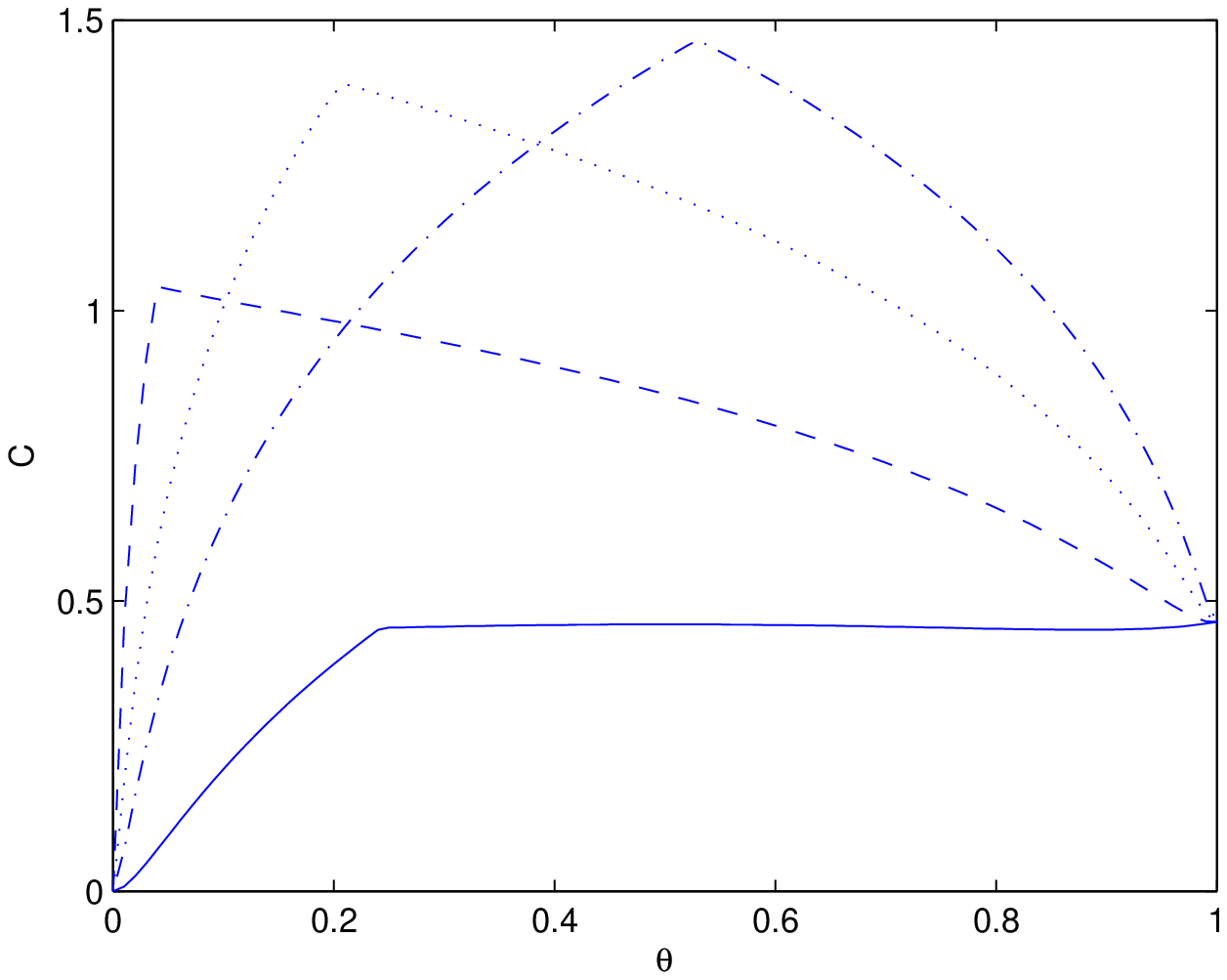}
\caption{Repetition coding  DF rate vs. $\theta$. $P =1$. (1) the
dashed line $\sigma_{sd}=1, \sigma_{sr}=10, \sigma_{rd}=2$; (2) the
dotted line $\sigma_{sd}=1, \sigma_{sr}=6, \sigma_{rd}=3$; (3) the
dashdot line $\sigma_{sd}=1, \sigma_{sr}=4, \sigma_{rd}=4$; (4) the
solid line $\sigma_{sd}=1, \sigma_{sr}=2, \sigma_{rd}=1$; %(1) and
%(4) are partly overlapped because they have the same s-d and r-d
%channel conditions.
} \label{fig:6}
\end{center}
\end{figure}

\begin{figure}
\begin{center}
\includegraphics[width = 0.43\textwidth]{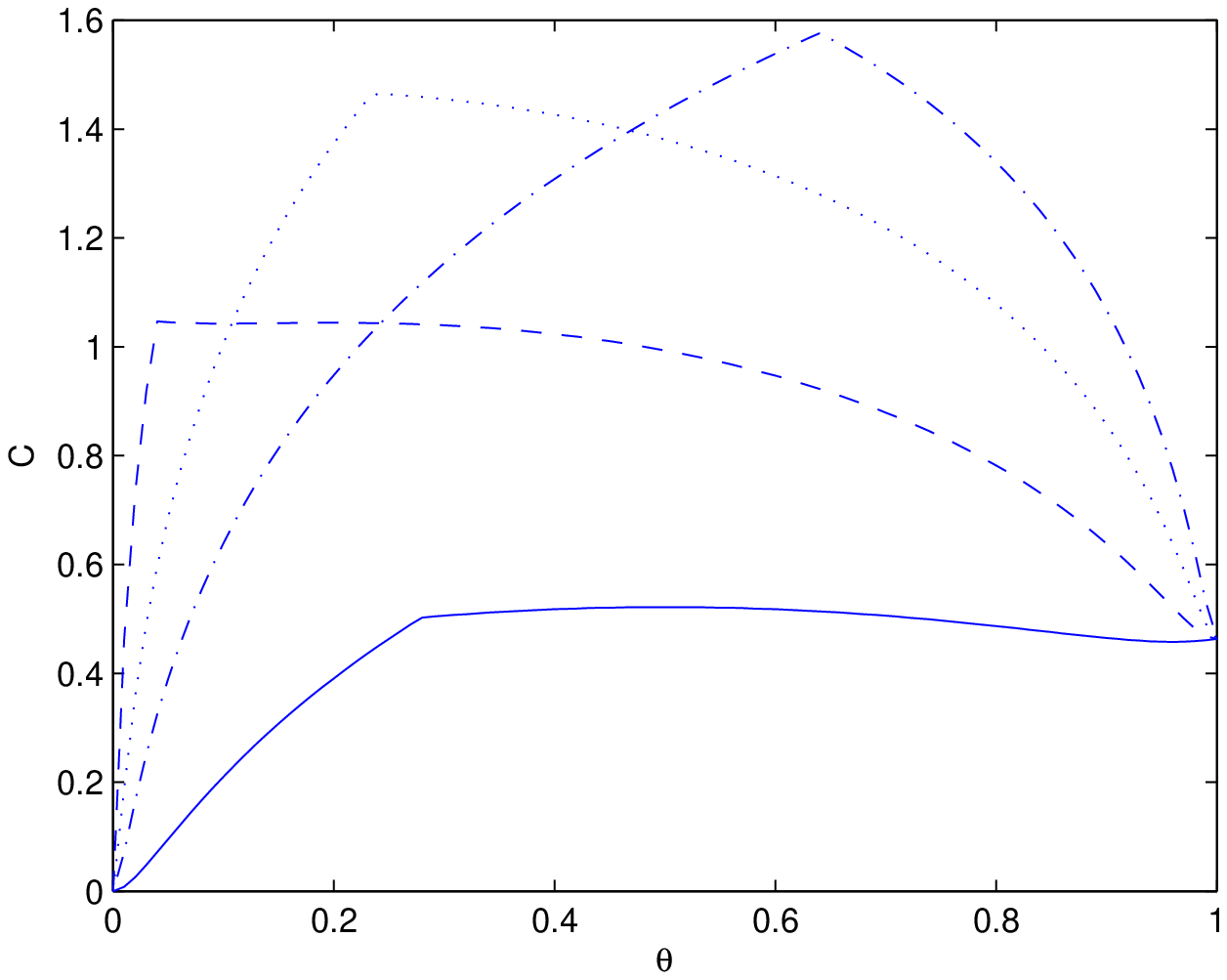}
\caption{Parallel coding DF rate vs. $\theta$. $P =1$.(1) the dashed
line $\sigma_{sd}=1, \sigma_{sr}=10, \sigma_{rd}=2$; (2) the dotted
line $\sigma_{sd}=1, \sigma_{sr}=6, \sigma_{rd}=3$; (3) the dashdot
line $\sigma_{sd}=1, \sigma_{sr}=4, \sigma_{rd}=4$;(4) the solid
line
$\sigma_{sd}=1, \sigma_{sr}=2, \sigma_{rd}=1$. %(1)and (4) are partly
%overlapped because they have the same s-d and r-d channel
%conditions.
} \label{fig:7}
\end{center}
\end{figure}

We consider in this paper that there is a cost associated with
cooperation. This cost is the power and time dedicated to learn
relay-destination channel. This cost is more pronounced in the
presence of a low total power constraint. Figs. \ref{fig:5},
\ref{fig:6}, and \ref{fig:7} plot the achievable rates when $P=1$.
We can see that DF have a better performance than AF at low power
levels. Generally, cooperation gives more gains in the low power
regime. However, as indicated by the solid-lined curves, if the
quality of the source-relay and relay-destination channels is
comparable to that of the source-destination channel, there is
little or no gain through cooperation.

\end{document}